\documentclass{iopart}
\usepackage{iopams}
\begin{document}

\title{Time-dependent fields of a current-carrying wire}
\author{D V Red\v zi\' c$^1$ and V Hnizdo$^2$}

\address{$^1$Faculty of Physics, University of Belgrade, PO
Box 44, 11000 Beograd, Serbia}
\address{$^2$National Institute for Occupational Safety and Health,
Morgantown, WV 26505, USA} \eads{\mailto{redzic@ff.bg.ac.rs}}

\begin{abstract}
The electric and magnetic  fields of an infinite straight wire
carrying a steady current which is turned on abruptly are determined
using Jefimenko's equations, starting from the standard assumption
that the wire is electrically neutral in its rest frame. Some
nontrivial aspects of the solution are discussed in detail.
\end{abstract}

\section{Introduction}
Consider an infinite straight linear wire carrying the current $I(t)
= 0$ for $t \leq 0$, and $I(t) = I_0$ for $t > 0$. That is, a
constant current $I_0$ is turned on abruptly at time $t = 0$. What
are the resulting electric and magnetic fields? This apparently
simple electrodynamic problem is posed and solved as Example 10.2 in
the excellent textbook of Griffiths \cite{DJG}. Starting from the
standard assumption that the wire is electrically neutral in its
rest frame, without or with the current, which implies that the
scalar potential $V$ is zero, the retarded vector potential $\bi A$
is calculated, and then the electric and magnetic fields are
obtained according to $\bi E = -\partial \bi A/ \partial t$ and $\bi
B = \bnabla \times\bi A$, respectively. (Note that in this case the
Coulomb and Lorenz gauges lead to same potentials since the wire is
electrically neutral.) While this solution is correct, we believe
that the problem has some intriguing aspects  and as such deserves
further attention. In this note we present a solution to the problem
using Jefimenko's equations and point out some pitfalls which could
be dangerous for novices. Thus, hopefully, our analysis should be
instructive for advanced undergraduate and beginning graduate
students.

\section{Solution using retarded potentials}
For the convenience of the reader, we first give the solution using
retarded potentials, in some more detail than that given in
Griffiths's book.

As is well known, the retarded vector potential $\bi A(\bi r,t)$ at
field point $\bi r$ and  time $t$ is given by
\begin {equation}
\bi A(\bi r,t) = \frac {\mu_0}{4\pi}\int \frac {\pmb J (\pmb
r',t_r)}{|\pmb r - \pmb r'|}\,\rmd^3r',
\end {equation}
where $\bi J (\bi r',t_r)$ is the current density at a source point
$\pmb r'$ and the retarded time $t_r= t - |\bi r - \bi r'|/c$. Let
the infinitely long wire lie along the $z$ axis. The current in the
wire, which is assumed to have an infinitesimal cross section, is
turned on abruptly at $t = 0$, and thus the current density can be
expressed as
\begin{equation}
\bi J(s,t) = \frac{I_0}{2\pi} \frac{\delta(s)}{s}\Theta (t)\hat {\bi
z},
\end{equation}
where $s$ is the distance from the wire, $\delta(s)$ is the
one-dimensional Dirac delta function normalized as $\int_0^{\infty}
\delta(s)\,\rmd s =1$ and $\Theta(t)$ is the Heaviside step
function,
\begin{equation}
 \Theta(t) =\left\{
  \begin{array}{lll}
  0,& {\rm if} & t \leq 0 \\
  1, & {\rm if} & t > 0.
  \end{array}
  \right.
\end{equation}
The setup is not realistic, but, in principle, it could be realized
approximately with a large superconducting loop of negligible cross
section in an inhomogeneous axially symmetric magnetic field, the
symmetry axis coinciding with the axis of the loop. If the loop,
initially at rest {\it with no current}, is moved quickly along the
symmetry axis into a new resting position, a persistent current is
produced in it, since the total magnetic flux through the loop is
constant (cf, e.g., \cite{RHR}).

Equations (1) and (2) imply that the vector potential $\bi A$ at a
distance $s$ from the wire is given by
\begin {eqnarray}
\bi A(s,t) &= \frac{\mu_0}{4\pi}\,\hat {\bi z}\int_0^{\infty}
s'\,\rmd s' \int_0^{2\pi}\rmd\phi'\int_{-\infty}^{\infty}\rmd z'
\frac{I_0}{2\pi} \frac{\delta(s')}{s'}\frac{\Theta
(t -d/c)}{d} \nonumber \\
&=\frac{\mu_0 I_0}{4\pi}\,\hat {\bi z}\int_{-\infty}^{\infty} \frac
{\Theta(t - \sqrt{s^2 + z'^2}/c)}{\sqrt{s^2 + z'^2}}\,\rmd z'.
\end {eqnarray}
Here, cylindrical coordinates $s,\phi,z$ are used and $d{=}[s^2{+}s'^2
{-}2s s'\!\cos(\phi{-}\phi') {+}(z{-}z')^2]^{\!\frac{1}{2}}$, which is the
distance between the field point $(s,\phi,z)$ and a source point
$(s',\phi',z')$; in the second line, the integration with respect to
$s'$ and a transformation $z-z' \to z'$ reduce the distance to
$(s^2+z'^2)^{1/2}$. As demanded by the problem's symmetry, the
vector potential is independent of $z$ and $\phi$. The
Heaviside-function factor in the integrand of the integral in the
second line of (4) causes the potential to vanish at times $t<s/c$
and limits the integration interval to the values of $z'$ satisfying
\begin {equation}
|z'| \leq \sqrt{c^2t^2 - s^2}.
\end {equation}
Integral (4) for the vector potential $\bi A$ thus evaluates
as\footnote {The standard convention is understood according to
which $f(x)\Theta(x-x_0) =0$ whenever $x\le x_0$, even when the
expression $f(x)$ happens not to be defined at these values of $x$.}
\begin {eqnarray}
\bi A(s,t) &= \frac{\mu_0 I_0}{4\pi}\,\hat{\bi z}\Theta(t-s/c)
\int^{\sqrt{c^2t^2 - s^2}}_{-\sqrt{c^2t^2 - s^2}} \frac {\rmd
z'}{\sqrt{s^2
 +z'^2}}\nonumber \\
&= \frac{\mu_0 I_0}{2\pi}\,\hat{\bi z} [\ln(ct + \sqrt{c^2t^2-
s^2})-\ln s] \Theta(t-s/c).
\end{eqnarray}

The electric field is therefore given by
\begin {eqnarray}
\bi E(s,t) &= - \partial \bi A(s,t)/\partial t \nonumber\\
&= -\frac{\mu_0 I_0}{2\pi}\,\hat{\bi z}\frac{c}{\sqrt{c^2t^2 -
s^2}}\, \Theta(t- s/c)
\end{eqnarray}
and the magnetic field by\footnote {It is perhaps worthwhile to note
that expression (6) for the vector potential $\bi A(s,t)$ resembles the quasi-static vector potential at a distance $s$ from the midpoint of a straight wire of finite length $2l=2\sqrt{c^2t^2 - s^2}$ carrying a constant current $I_0$. A calculation of the magnetic field according to $\bi B = \bnabla \times \bi A$ in which the distance-dependent length $2l$ is treated as a constant would be equivalent to the use of the Biot--Savart law. However, this law is not applicable beyond the quasi-static regime (cf., e.g., \cite{VH2003}).}
\begin {eqnarray}
\bi B(s,t) &= \bnabla\times\bi A(s,t) = -(\partial
A_z/\partial{s})\,\hat{\!\bphi}
\nonumber \\
&= \frac{\mu_0 I_0}{2\pi s}\,\hat{\!\bphi}\frac{ct}{\sqrt{c^2t^2
-s^2}\, }\Theta(t - s/c).
\end {eqnarray}
In both (7) and (8), the delta-function terms that arose from the
derivatives of the Heaviside step function in (6) dropped out on
account of the property
\begin {equation}
f(x)\delta(x - x_0) = f(x_0)\delta(x - x_0)
\end {equation}
of the delta function. Inspecting equations (7) and (8) we see that
the fields $\bi E(s,t)$ and $\bi B(s,t)$ attain in the limit
$t\to\infty$ their familiar static values 0 and $(\mu_0 I_0/2\pi
s)\,\hat{\!\bphi}$, respectively, and that both these fields diverge
when $t \to s/c$.

\section{Solution using Jefimenko's equations}
As is now well known, starting from the retarded solution to the
inhomogeneous wave equations for the fields $\bi E$ and $\bi B$
\cite{OJ,JDJ}, or from the familiar retarded potentials
\cite{DJG,GH}, or otherwise \cite{TCT}, the time-dependent
generalizations of the Coulomb and Biot-Savart laws can be derived:
\begin{equation}
\bi E (\bi r, t) = \frac {1}{4\pi \epsilon_0}\int \left[ \frac
{\varrho (\bi r', t_r)}{{\cal R}^3}\,\pmb{\cal R} +  \frac {\dot
{\varrho} (\bi r', t_r)}{c{\cal R}^2}\,\pmb {\cal R} - \frac {\dot
{\bi J} (\bi r', t_r)}{c^2{\cal R}}\right ]d^3r',
\end {equation}
\begin {equation}
\bi B (\bi r, t) = \frac {\mu_0}{4\pi }\int \left[ \frac {\bi J (\bi
r', t_r)}{{\cal R}^3} + \frac {\dot {\bi J} (\bi r', t_r)}{c{\cal
R}^2}\right ]\times \pmb {\cal R}\, d^3r',
\end {equation}
where $\varrho$ is the volume charge density, $\pmb {\cal R} \equiv
\bi r - \bi r'$, and the dots denote partial differentiation with
respect to time. These equations, showing explicitly {\it true}
sources of $\bi E$ and $\bi B$, were first derived by Jefimenko
\cite{OJ}. We now shall calculate the fields $\bi E$ and $\bi B$ in
the problem at hand using Jefimenko's equations.

In our problem, the charge density vanishes,
\begin {equation}
\varrho = 0,
\end {equation}
since by assumption the wire is electrically neutral, and using
equation (2) we get
\begin{equation}
\bi J(s',t_r) = \frac{I_0}{2\pi} \frac{\delta(s')}{s'}\,\Theta (t -
\sqrt{s^2 + z'^2}/c)\,\hat {\bi z},
\end{equation}
\begin{equation}
\dot {\bi J}(s',t_r) = \frac{I_0}{2\pi}\,\Theta(t-s/c)
\frac{\delta(s')}{s'}\,\delta (t - \sqrt{s^2 + z'^2}/c)\,\hat{\bi
z}.
\end{equation}
Here, cylindrical coordinates are used again, and the delta-function
property (9) and the taking, with no loss of generality, the field
coordinate $z$ to be 0 reduced the retarded time $t_r$ to the same
value as that in equation (4); the step-function factor in (14)
expresses the fact that the step function in (13) entails that not
only the current density itself but also its partial time derivative
vanishes for times $t<s/c$. Substitution into Jefimenko's equation
(10) then gives
\begin{equation}
\fl \bi E(s,t) = -\frac{I_0}{4\pi \epsilon_0 c^2}\,\hat {\bi z}\,
\Theta(t-s/c)\int_0^{\infty}s'\,\rmd s'\,\frac{\delta(s')}{s'}
\int_{-\infty}^{\infty}\rmd z'\frac{\delta (t - \sqrt{s^2 +
z'^2}/c)} {\sqrt{s^2 + z'^2}},
\end{equation}
which can be  evaluated easily  using the decomposition of the delta
function \cite{JDJ,VSV}
\begin{equation}
\fl \delta(t - \sqrt{s^2 + z'^2}/c) = \frac{c^2t}{\sqrt{c^2t^2 -
s^2}}\left [\delta (z' - \sqrt{c^2t^2 - s^2}) + \delta (z' +
\sqrt{c^2t^2 - s^2})\right ]
\end{equation}
as
\begin{equation}
\bi E(s,t) = -\frac{\mu_0 I_0}{2\pi }\hat {\bi z}\,
\frac{c}{\sqrt{c^2t^2 - s^2}}\,\Theta(t-s/c),
\end{equation}
in full agreement with the electric field (7), obtained using the
retarded vector potential. In a similar fashion, using equations
(11), (13), (14) and (16) we obtain
\begin{eqnarray}
\fl \bi B(s,t) &= \frac{\mu_0
I_0}{4\pi}\,\Theta(t{-}s/c)s\,\hat{\!\bphi} \left[\int_{-\sqrt{c^2t^2
-s^2}}^{\sqrt{c^2t^2 -s^2}} \frac{\rmd z'}{(s^2 +
z'^2)^{3/2}}+\frac{1}{c}\int_{-\infty}^{\infty} \frac {\delta (t {-}
\sqrt{s^2 + z'^2}/c)}{s^2 + z'^2}\,\rmd z'\right]
\nonumber \\
\fl &= \frac{\mu_0 I_0}{2\pi s}\,\hat {\!\bphi}\,\frac{ct}
{\sqrt{c^2t^2 -s^2}}\,\Theta(t-s/c),
\end{eqnarray}
in full agreement with the magnetic field (8), obtained using the
retarded vector potential.

\section{Discussion}
At first sight, the fact that the fields $\bi E$ and $\bi B$
obtained diverge when $t \to s/c+$ while vanishing for $t < s/c$ may
seem disturbing. However, a closer examination reveals that, in the
correct solution to the problem, $\bi E$ and $\bi B$ {\it must} tend
to infinity when $t \rightarrow s/c+$. This is more transparent
through the use of Jefimenko's equations (10) and (11) than the use
of a retarded vector potential. It is clear from equations (13),
(14) and (16) that the abrupt turning of the current on at $t = 0$
necessarily yields an infinite time derivative of the current
density, producing in the fields a cylindrical `shock wave' that
diverges at the time $t=s/c$ at a distance $s$ from the wire.
Similar to the instructive example of Jackson of an abruptly turned
on electric dipole \cite{JDJ2}, the diverging fields $\bi E$ and
$\bi B$ are here artifacts of the unphysical, instantaneous turn-on
of the current.\footnote {We remind the reader that a similar
situation is found in the well-known $RC$-circuit problem of
charging a capacitor  of capacitance $C$ by connecting it
instantaneously to a constant voltage $V$ through a resistor of
resistance $R$, assuming that the charge $Q$ on the positive plate
is zero at $t = 0$. The standard (tacit) assumption that the
inductance $L$ of the circuit is zero then leads to the equation
$V=Q/C + R I$. In the unphysical  setup of the problem (the abrupt
closing of a circuit with $L = 0$), the correct solution {\it must}
satisfy the unphysical initial condition $I(t = 0) = V/R$, despite
the fact that the current $I$ vanishes for $t < 0$.}

The divergences disappear if the current is not turned on abruptly
but is increased gradually during a short time interval $\tau$. As a
simple example, assume that the current increases linearly from zero
at $t = 0$ to a steady non-zero value $I_0$ at $t \ge \tau$,
replacing accordingly expression (2) for the current density by
\begin{equation}
\bi J(s,t) = \frac{I_0}{2\pi} \frac{\delta(s)}{s}\left[
\frac{t}{\tau}\,\Theta (t) - \frac{t-\tau}{\tau}\,\Theta(t - \tau)
\right ]\hat {\bi z}.
\end{equation}
Using Jefimenko's equations, the resulting fields are then obtained
to be
\begin{eqnarray}
\fl\bi E_{\tau}(s,t)&=-\frac{\mu_0I_0}{2\pi}\,\frac{\hat{\bi z}}{\tau}\left[ \ln\left(ct/s +\sqrt{c^2t^2-s^2}/s\right)\Theta(t-s/c)\right.\nonumber\\
\fl&\quad\left.-\ln\left(c(t-\tau)/s
+\sqrt{c^2(t-\tau)^2-s^2}/s\right)\Theta(t-\tau -s/c)\right]
\label{Etau}
\end{eqnarray}
and
\begin{equation}
\fl \bi B_{\tau}(s,t)=\frac{\mu_0I_0}{2\pi c}\,
\frac{\hat{\!\bphi}}{s\tau}\left[\sqrt{c^2t^2-s^2}\Theta(t-s/c)
-\sqrt{c^2(t-\tau)^2-s^2}\Theta(t-\tau -s/c)\right]. \label{Btau}
\end{equation}
While the fields (\ref{Etau}) and (\ref{Btau}) are finite for any
non-zero parameter $\tau$, their limits $\tau\to 0$ can be shown
easily to be the fields (7) and (8), respectively, that diverge when
$t\to s/c$.

There is another query. The retarded vector potential and
Jefimenko's equations are both derived under the assumption that the
sources (charges and currents) are localized in a finite region of
space, but our problem involves an infinitely long current-carrying
wire.  Therefore, the question arises as to the validity of the
solution found.\footnote{Recall that in electrostatics the standard
solution to the Poisson equation is not generally valid for charge
distributions extending to infinity (cf., e.g., \cite{DVR,JCR}).}
However, inspecting equation (6) we see that for any given finite
$s$ and $t$ only a finite segment of the wire contributes to the
retarded vector potential. Figuratively speaking, retardation makes
the infinitely long  wire finite. Still, one should check that the
vector potential (6) satisfies the requisite inhomogeneous wave
equation,
\begin{equation}
\nabla^2 \bi A(s,t) - \frac{1}{c^2}\frac{\partial^2 \bi
A(s,t)}{\partial t^2} = -\frac{\mu_0 I_0}{2\pi}
\frac{\delta(s)}{s}\Theta (t)\,\hat {\bi z}, \label{inhomweq}
\end{equation}
and that the fields (7) and (8) satisfy all Maxwell's equations.

Let us check first whether Maxwell's equations are satisfied.
Straightforward calculations yield that the fields (7) and (8) are
divergenceless, confirming the equations $\bnabla \bdot{\bi
E}=\varrho/\epsilon_0$, where $\varrho{=}0$, and $\bnabla \bdot{\bi
B} =0$. While a straightforward calculation of the curl of the
electric field (7)  confirms that the fields obtained satisfy
Faraday's law,  a  similar calculation of the curl of the magnetic
field  (8)  using the standard cylindrical-coordinate formula
\begin{equation}
\bnabla \times F(s)\,\hat{\!\bphi}=\frac{1}{s}\,\frac{\partial(s
F)}{\partial s}\, \hat{\bi z} \label{cylnabla}
\end{equation}
appears to yield only that $\bnabla\times \bi B = \epsilon_0\mu_0
\partial \bi E/\partial t$, instead of the full Amp\`ere--Maxwell
law, which reads in our case
\begin{equation}
\bnabla \times \bi B = \frac{\mu_0 I_0}{2\pi}\,
\frac{\delta(s)}{s}\,\Theta (t)\hat {\bi z}+ \epsilon_0
\mu_0\frac{\partial \bi E}{\partial t}. \label{Ampere-Maxwell}
\end{equation}
Here, however, it is important to bear in mind that when the sources
of a field are idealized point, line or surface distributions of
charge and/or current, described by generalized functions such as
the Dirac delta function, great care must be taken to employ in the
field or potential differential equations generalized
(distributional) derivatives instead of the usual (classical)
ones\footnote{For example, calculating the Laplacian of the
potential $1/r$ of a unit point charge using classical derivatives
yields $\nabla^2(1/r) = 0$ for all $r > 0$; at  $r = 0$, the
classical Laplacian is simply not defined. In the well-known
relation $\nabla^2(1/r) = - 4\pi\delta (\bi r)$, the Laplacian is in
fact the generalized one, as it must be since this relation
expresses equality of two generalized functions. To avoid confusion,
some authors denote generalized differential operators by a bar,
writing thus $\bar{\nabla}^2(1/r) = - 4\pi\delta (\bi r)$
\cite{Kanwal,Estrada,VH}.}. Keeping this in mind, a careful
examination of the differential operation on the magnetic field (8)
implied by formula (\ref{cylnabla}) reveals that it involves an
expression that can be written as the Laplacian of the natural
logarithm of the cylindrical coordinate $s$,\footnote{This fact is
perhaps more transparent when using $\bnabla \times \bi B =
\bnabla(\bnabla \bdot \bi A) - \nabla^2\bi A$ and recognizing that
for $\bi A$ expressed by equation (6), $\nabla^2(\ln s)$ appears
explicitly in the calculation of $\nabla^2\bi A$.}
\begin{equation}
\frac{1}{s} \frac{\partial }{\partial s}\left [s \left
(\frac{1}{s}\right )\right ] =\frac{1}{s} \frac{\partial }{\partial
s}\left [s \frac{\partial }{\partial s}(\ln s)\right ] =
\nabla^2(\ln s). \label{Lapln}
\end {equation}
In terms of the usual (classical) derivatives, this Laplacian
vanishes for all $s>0$ and is not defined at $s=0$, but in view of
the fact that the current density involves the delta function we
should use here the relation
\begin{equation}
\nabla^2(\ln s)=\frac{\delta(s)}{s}. \label{genLapln}
\end{equation}
Employing this relation  in the evaluation of $\bnabla \times \bi
B$, it is found easily that the fields (7) and (8) now satisfy the
full Amp\`ere--Maxwell law (\ref{Ampere-Maxwell}).

Since the relation (\ref{genLapln}) may appear to be novel to some
readers, we give an informal proof of it using a limiting procedure
in which $\ln s$ is regularized as $\ln\sqrt{s^2+a^2}$ and the limit
$a\to 0$ is taken after integrating the product of
$\nabla^2\ln\sqrt{s^2+a^2}$ and a well-behaved `test' function
$f(s)$:
\begin{equation}
\fl \lim_{a\to 0}\int_0^\infty\nabla^2\ln\sqrt{s^2+a^2}f(s)s\,\rmd s
=\lim_{a\to
0}\int_0^\infty\left[\frac{2a^2}{(s^2+a^2)^2}\right]f(s)s \,\rmd s.
\label{limit}
\end{equation}
The expression in square brackets is the Laplacian of
$\ln\sqrt{s^2+a^2}$, which is now a well-behaved function of $s$ for
any $a\ne 0$; its  integral over the whole plane is independent of
$a$, equalling  $2\pi$. Splitting the integral into integrals  over
intervals $0\le s \le S$ and $S\le s <\infty$ so that the function
$f(s)$ can be expanded in the first interval in a Taylor series
around $s=0$, the limit (\ref{limit}) can be evaluated as
\begin{eqnarray}
\fl &\lim_{a\to 0}\int_0^S \frac{2a^2}{(s^2+a^2)^2}\sum_{n=0}^\infty
\frac{f^{(n)}(0)s^n}{n!}\,s\,\rmd s +\lim_{a\to
0}\left[2a^2\int_S^\infty\frac{f(s)}{(s^2+a^2)^2}\, s
 \,\rmd s\right] \nonumber \\
\fl &\quad=\lim_{a\to 0}\left[\frac{f(0)}{1{+}a^2/S^2}+
\left(a\,{\rm arctg}\frac{S}{a}-\frac{S
a^2}{S^2{+}a^2}\right)\!f'(0) +O(a^2,a^2\ln a)+\cdots\right]=f(0).
\label{limit2}
\end{eqnarray}
The second limit in the first line vanished since the integral of
$f(s)s/(s^2+a^2)^2$, where $f(s)$ is assumed to be integrable over
the whole interval $(0,\infty)$, is guaranteed to converge  to a
value  that remains finite when $a\to 0$. Having thus shown that
\begin{equation}
\lim_{a\to 0}\int_0^\infty\nabla^2\ln\sqrt{s^2+a^2}f(s)s\,\rmd s
=f(0),
\end{equation}
where $f(s)$ may be any well-behaved function of $s$, we can write
\begin{equation}
\lim_{a\to 0}\nabla^2\ln\sqrt{s^2+a^2} =\delta(s)/s, \label{limit3}
\end{equation}
which is the relation (\ref{genLapln}) with  $\nabla^2(\ln
s)\equiv\lim_{a\to 0}\nabla^2\ln\sqrt{s^2+a^2}$.\footnote{Strictly
speaking, the limit here and in (\ref{limit3}) is the weak limit
(cf., e.g., \cite{VSV}) and the Laplacian in (\ref{genLapln}) is the
generalized (distributional) one (cf., e.g., \cite{Kanwal}).} (Note
that the relation (\ref{genLapln}) may be interpreted as the Poisson
equation  for the electrostatic potential $-(1/2\pi \epsilon_0)\ln
s$ of a charge density $\delta(s)/2\pi s$, which is that of an
infinitely long straight line of charge of unit line charge density
\cite{JDJ,DVR,JCR,JAS}.)

Checking whether the vector potential (6) satisfies the
inhomogeneous wave equation (\ref{inhomweq}) is a somewhat
cumbersome but in every step straightforward calculation. Using the
relation (\ref{genLapln}) and making  extensive use of relation (9)
in the terms containing $\delta(t-s/c)$, which arise through
differentiations of $\Theta(t-s/c)$, we obtain
\begin{equation}
\nabla^2 \bi A(s,t) - \frac{1}{c^2}\frac{\partial^2 \bi
A(s,t)}{\partial t^2} = -\frac{\mu_0 I_0}{2\pi}
\frac{\delta(s)}{s}\Theta (t-s/c)\,\hat {\bi z},
\end{equation}
which confirms, as it must\footnote{The facts that  $\bnabla
\times\bi B =\mu_0 \bi J + \epsilon_0\mu_0 \partial \bi E/\partial
t$, $\bi B = \bnabla \times \bi A$, $\bi E = - \partial \bi
A/\partial t$ and $\bnabla \bdot \bi A = 0$  already imply that
$\Box \bi A = - \mu_0 \bi J$.}, that the vector potential (6)
satisfies the wave equation (\ref{inhomweq}) since, on account of
(9), $\delta(s)\Theta(t{-}s/c) = \delta(s)\Theta(t)$.

Closing the discussion, we remark that the solution to the problem
of finding the fields due to an infinite straight linear wire
carrying a constant current that is turned on abruptly appears to be
relevant to the related problem of finding charges and fields in a
current-carrying wire of finite cross section, where it appears that
an infinite time is needed for establishing a stationary charge
distribution  \cite{DVR2}. However, it should be noted that the
time-dependent fields obtained here approach their values at $t\to
\infty$ very rapidly. For example, the field (18) at a distance
$s=1$ m and a time $t =3$ $\mu$s differs from its asymptotic value
by less than 1 part in a million (the wire is then required to be at
least 2 km long).

\section{Conclusions}

We calculated the electric and magnetic fields of an infinitely long
wire carrying a constant current that is turned on abruptly, using
Jefimenko's equations.  Our calculations confirmed the terse
solution given to this problem in an example of Griffiths's text,
but our method appears to provide more insight than the standard
approach via the retarded potentials: the divergence of the
resulting fields $\bi E(s,t)$ and $\bi B(s,t)$ when $t \rightarrow
s/c$ then can be seen easily as a necessary consequence of the
`unphysical' setup of the problem. We also calculated the fields for
a more realistic case in which the current in the wire is increased
linearly in a nonzero time to its final steady value. We believe
that our analysis was an instructive demonstration of not only the
power but also the possible pitfalls of using the delta function in
the calculations of electrodynamics.

\section*{Acknowledgments}
We thank David Griffiths for illuminating correspondence and the anonymous referee for constructive comments. DVR
acknowledges support of the Ministry of Science and Education of the
Republic of Serbia, project No.\ 171028. VH co-authored this paper in
his private capacity; no official support or endorsement by the
Centers for Disease Control and Prevention is intended or should be
inferred.


\Bibliography{99}
\bibitem{DJG} Griffiths D J 1999 {\em Introduction to Electrodynamics}
     3rd edn (Upper Saddle River, NJ: Prentice Hall)
\bibitem {RHR} Romer R H 1990 The motion of a superconducting loop
     in an homogeneous magnetic field: the harmonic oscillator equation
     in an unfamiliar setting {\it Eur.
      J. Phys.} {\bf 11} 103--6
\bibitem{VH2003} Hnizdo V 2003 Comment on `About the magnetic field of a finite wire' {\it Eur. J. Phys.} {\bf 24} L15--L16
\bibitem{OJ} Jefimenko O D 1989 {\it Electricity and Magnetism} 2nd
      edn (Star City, WV: Electret Scientific)
\bibitem{JDJ} Jackson J D 1999 {\it Classical Electrodynamics} 3rd
      edn (New York: Wiley)
\bibitem{GH} Griffiths D J and Heald M A 1991 Time-dependent
      generalizations of the Biot-Savart and Coulomb laws
      {\it Am. J. Phys.} {\bf 59} 111--7
\bibitem{TCT} Ton T C 1991 On the time-dependent, generalized
      Coulomb and Biot-Savart laws {\it Am. J. Phys.} {\bf 59} 520--8
\bibitem{VSV} Vladimirov V S 1979 {\it Generalized Functions in
      Mathematical Physics} (Moscow: Mir)
\bibitem{JDJ2} Jackson J D 2010 Comment on `Maxwell equations and the
      redundant gauge degree of freedom' {\it Eur. J. Phys.} {\bf 31}
      L79--L84
\bibitem {DVR} Red\v zi\' c D V 2012 The calculation of the
      electrostatic potential of infinite charge distributions {\it Eur.
      J. Phys.} {\bf 33} 941--6
\bibitem{JCR} Jim\' enez J L, Campos I and Roa-Neri J A E 2012
      Confusing aspects in the calculation of the electrostatic potential
      function of an infinite line of charge {\it Eur. J. Phys.} {\bf 33}
      467--71
\bibitem{Kanwal}  Kanwal R P 2004 {\em Generalized Functions: Theory and
     Applications} 3rd edn (Boston: Brikh{\"a}user)
\bibitem{Estrada} Estrada R and Kanwal R P 1995 The appearance of nonclassical terms in the analysis of point-source fields {\it Am. J. Phys.} {\bf 63} 278--78
\bibitem{VH} Hnizdo V  2011 Generalized second-order partial derivatives
      of $1/r$ {\it Eur. J. Phys.} {\bf 32} 287--97
\bibitem{DVR2} Red\v zi\'c D V 2012 Charges and fields in a current-carrying
      wire {\it Eur. J. Phys.} {\bf 33} 513--23
\bibitem{JAS} Stratton J A 1941 {\it Electromagnetic Theory} (New
       York: McGraw-Hill)
\endbib
\end{document}